\documentclass[twocolumn,showpacs,preprintnumbers,amsmath,amssymb,aps]{revtex4}
\usepackage{graphicx}
\begin{document}

\title{Statics and Dynamics of Yukawa Cluster Crystals on Ordered Substrates 
} 
\author{C. Reichhardt and C. J. Olson Reichhardt} 
\affiliation{ 
Theoretical Division, Los Alamos National Laboratory, Los Alamos, New Mexico 87545
}

\date{\today}
\begin{abstract}
We examine the statics and dynamics of 
particles with repulsive Yukawa interactions
in the presence of a two-dimensional triangular substrate 
for fillings of up to twelve
particles per potential minimum.  
We term the ordered states Yukawa cluster crystals and show that
they are distinct from the
colloidal molecular crystal states found at low fillings.
As a function of substrate and interaction strength at fixed particle
density we find a series of novel crystalline states
that we characterize using the structure factor. 
For fillings greater than four,
shell and ring structures form at each 
potential minimum 
and can exhibit sample-wide orientational order.
A disordered state can appear between ordered states as
the substrate strength varies.
Under an external drive, 
the onsets of different orderings 
produce
clear changes in the critical depinning force, including 
a peak effect phenomenon that has generally only previously 
been observed in systems with random substrates. 
We also find a rich variety of
dynamic ordering transitions that can be observed 
via changes in the structure factor and features in the 
velocity-force curves. 
The dynamical states encompass a variety of  moving structures including
one-dimensional stripes, smectic ordering, 
polycrystalline states, triangular lattices, and symmetry locking states.
Despite the complexity of the system, 
we identify several generic features of the dynamical phase
transitions which we map out in a series of phase diagrams.
Our results have implications
for the structure and depinning of colloids on 
periodic substrates, vortices in superconductors
and Bose-Einstein condensates, Wigner crystals, and dusty plasmas. 
\end{abstract}
\maketitle

\vskip2pc

\section{Introduction}
The creation of new types of crystalline or partially ordered
states and the dynamics of assemblies of interacting particles 
have attracted much attention both in 
terms of the basic science of self-assembly and
dynamic pattern formation as well as
for applications utilizing self-assembly processes. 
One of the
most extensively studied systems exhibiting this behavior is
assemblies of colloidal particles, where the equilibrium structures can be 
tuned by changing the
directionality of the colloid-colloid interactions \cite{Self,M1}. 
Since it can be difficult to control and tune the exact 
form of the interaction, 
another approach is to use colloids with well defined interactions that are
placed on some type of ordered substrate. 
Optical trapping techniques are one of the most common methods of creating
periodic substrates for colloids \cite{Ashkin}.
Studies of colloidal ordering and melting  
for one dimensional (1D) periodic substrate arrays have revealed
ordered colloidal crystalline structures as well as smectic type structures
where the colloids are crystalline in one direction 
and liquidlike in the other 
\cite{Clark,J,Wei,Nelson,Bech1,Bech2,P,Stark,Colloid}. 
These experiments show
that the substrate strength strongly influences the type of
colloidal structure that forms and that as the substrate strength
increases, the resulting enhancement of fluctuations can
induce a transition from crystalline to smectic order
\cite{Clark,Bech1,Wei}.    
Numerous different colloidal crystalline structures can also appear on 
1D substrate arrays of fixed periodicity when the colloid density is varied
\cite{Nelson,Bech2}. 

More recent studies addressed colloidal
ordering on two-dimensional (2D) periodic substrates
\cite{Grier,Spalding,Korda,Reichhardt,Brunner}.  
In these systems, the filling factor $f$ is defined as the
number of colloids per substrate minimum.
For integer fillings $f=n$, the colloids in each minimum can form
an effective rigid $n$-mer, such as a dimer or trimer
\cite{Reichhardt,Brunner,Scalettar,Agra,Frey,Tr,Samir}. 
The $n$-mers have an orientational degree of freedom, and depending on
the effective interaction between neighboring $n$-mers, all the $n$-mers
may align into a ferromagnetically ordered state, sit perpendicularly to
their neighbors in an antiferromagnetically ordered state, or form other
orientationally ordered states.
For square and triangular substrate arrays, 
$n$-mer states have been studied up to $n = 4$ 
\cite{Reichhardt,Agra,Frey,Tr};
however, it is not known what structures would form at higher 
fillings when the simple
picture of rigid $n$-mers no longer applies.
Studies of the ordering of bidisperse colloidal assemblies with
two different charges on 2D periodic substrates produced novel ordered phases,
while a pattern switching could be induced by application of 
an external field \cite{Samir}. 
Similar pattern switching
also occurs for colloids with monodisperse charges under external
driving \cite{CO}.   
The colloidal $n$-mer states have been termed ``colloidal molecular crystals,'' 
and for conditions under which they loose their orientational ordering,
they are referred to as ``colloidal plastic crystals.''
Colloidal molecular crystals appear for integer fillings $f=n$.
At fractional fillings such as $f=3/2$, $5/2$, or $7/2$, it is possible 
for ordered composite states to form containing two coexisting species 
of $n$-mers; however, for other fractional fillings,
the system is frustrated and the $n$-mer states are disordered \cite{Olson2}.
Other studies have shown that novel orderings 
appear when the 2D substrate array has quasicrystalline order
\cite{Roth}.

Once colloidal crystals have formed on a substrate, the driven dynamics can
be explored by applying an
external driving field to the sample.
A variety of dynamical locking phases can occur 
in which the colloids preferentially flow along symmetry directions of the
substrate \cite{Korda2}. 
As the filling fraction $f$ is varied, 
a series of peaks in the critical force needed to 
depin the colloids occurs at integer values of $f$
indicating the existence of commensurability effects \cite{C}.
Recent experiments with strongly interacting colloids on two periodic
arrays show that kink-type dynamics can occur near $f=1$
\cite{Guth}.
It would be interesting to explore
higher fillings 
where new types of dynamics could emerge \cite{Bohlein}. 

Many of the same types of phenomena found 
in colloidal molecular crystals can also be realized
for other systems that can be modeled as interacting particles in the presence
of a 2D periodic substrate. 
For example, the antiferromagnetic ordering of dimer colloidal
molecular crystals on a square substrate was 
reproduced using vortices in Bose-Einstein condensates
confined by optical traps with two 
vortices per trap \cite{Pu} 
as well as with vortices trapped by large pinning sites
in type-II superconductors
\cite{Jensen,Yetis,Bending}. 
Experimental and numerical studies
of molecular ordering on periodic substrates show similar orderings 
\cite{Rohr}.  
Other systems where similar states could be realized include
classical electrons or dusty plasmas
with some form of substrate as well as
crystalline cold atoms on optical lattices.   

To our knowledge, previous studies of Yukawa 
interacting colloidal molecular crystals have focused only on systems with 
up to four colloids per trap. 
For the case of three colloids per trap, only a limited number of studies 
have considered the dynamics, 
and even in this limit there are several new features 
that we describe for the first time in this work.  
We show that at high fillings, 
the rigid $n$-mer picture breaks down and new cluster and ring states form.  
Several general features of the statics and dynamics 
emerge at these larger fillings which are independent of the specific filling. 

One of the key findings in our work is the development of 
orientationally ordered shell structures at higher fillings 
for the 2D arrays. 
The development of particle shell structures was studied previously
for repulsive particles in isolated individual traps, in systems
that include
classical charges \cite{Peeters1}, 
Wigner islands \cite{AK},
dusty plasmas \cite{Lin}, 
colloids \cite{Bubeck,Peeters2,Kong,Peeters3,Drocco}, 
charged balls in traps \cite{Even}, 
and vortices in confined geometries \cite{Misko,Zhao}. 
The shell ordering in these systems can be altered by the shape or
type of trap as well as by competing attractive and
repulsive interactions between the particles \cite{M2,Liu}. 
In our system, the particles within each shell can exhibit an 
additional ordering due to interactions 
with particles in the neighboring traps. 
We show that as a function of substrate strength, 
a rich variety of colloidal cluster crystals can be created 
depending on the filling, and that certain
shell structures are more stable than others. 
The number of particles in 
the shells and the number of shells depend on the substrate 
strength, and in certain
cases the particles form ring structures instead of shells. 
We also find
reentrant ordered phases as a function of substrate
strength.  For weak and strong substrates,
the system is ordered; 
however, for substrates of intermediate strength, the system is 
generally disordered.
We find that 
the depinning threshold can show distinct changes at the boundaries
between these different phases as a function of 
substrate strength, and remarkably, we find that  
in some cases it is possible for the depinning force to 
decrease with increasing substrate strength.
The velocity-force
curves contain clear signatures of the 
different phases corresponding to different modes of depinning.  

\section{Simulation}
We consider a 2D system with periodic boundary conditions in the
$x$ and $y$ directions. 
The sample contains $N_{c}$ particles 
interacting with a triangular sinusoidal substrate 
containing $N_{s}$ potential
minima. 
We focus on the case where the number of colloids $N_{c}$ is an 
integer multiple of $N_s$ such that $f=N_{c}/N_{s} = n$, 
where $f$ is the filling factor and $n$ is an integer.
A single particle $i$ responds to forces according to the
following overdamped equation of motion: 
\begin{equation}
\frac{d {\bf R}}{dt} = {\bf F}^{cc}_{i} + {\bf F}^{s}_{i}  + {\bf F}^{ext}  + {\bf F}^{T}_{i}.
\end{equation} 
The particle-particle interaction force 
${\bf F}^{cc}_{i} =-\sum^{N_{c}}_{j\neq i}\nabla V(R_{ij})$ and
the particle interactions are of a Yukawa form,
$V(R_{ij}) = E_{0}\exp(-\kappa R_{ij})/R_{ij}$, where 
$E_{0} = Z^{2*}/4\pi\epsilon\epsilon_{0}a_{0}$, 
$\epsilon$ is the solvent dielectric constant, 
$Z^{*}$ is the effective charge, $1/\kappa$ is the screening length,
$a_{0}$ is the unit of length which is of order a micron, 
and $R_{ij} = |{\bf R}_{i} - {\bf R}_{j}|$.
We model the substrate as a triangular lattice, 
\begin{equation}
{\bf F}^{s}_{i} = \sum^{3}_{k=1}A\sin\left(\frac{2\pi b_{k}}{a_{0}}\right)[\cos(\theta_{k}){\bf \hat x} - \sin(\theta_{k}){\bf \hat y}] 
\end{equation}
where $b_{k} = x\cos(\theta_{k}) - y\sin(\theta_{k}) + a_{0}/2$, 
$\theta_{1} = \pi/6$, $\theta_{2} = \pi/2$, and
$\theta_{3} = 5\pi/6$. $A$ is the amplitude of the substrate potential.
The initial colloid positions are obtained by simulated annealing,
where the colloids start in a high temperature molten state and 
are gradually cooled to a finite temperature or to $T = 0$. 
We have also considered other initialization procedures,
including placing a commensurate number 
of colloids in each substrate minima and letting the system relax. 
This procedure gives very similar initial states as the simulated annealing,
especially for strong substrates; 
however, for weaker substrates the configuration obtained through
simulated annealing is generally more disordered, 
as we discuss in more detail later.  
After annealing, we investigate the transport properties 
by applying an external drive 
${\bf F}^{ext} = F_{D}{\bf {\hat x}}$.
We increment $F_D$ from $F_D=0$
to some final value in steps of $\delta F_{D}$, 
and we wait $10^6$ simulation
time steps between increments to avoid transient effects. 
The depinning threshold is obtained
by measuring the average colloid velocity 
$V = N^{-1}_{c}\sum^{N_{c}}_{i} (d {\bf R}_{i}/dt) \cdot {\bf {\hat x}}$.
The value of $\delta F_{D}$ is modified according to the 
strength of the substrate. 
For weaker substrates we use smaller values 
of $\delta F_D$ in order to obtain an accurate depinning threshold.  

\section{Configurations and Dynamics for $f =$ 3, 4, and 5} 

We first consider the cases 
$f= 3$, 4, and $5$.  
Earlier numerical studies addressed the statics and dynamics for 
$f = 2$ and $3$ and
showed that for strong triangular substrates, 
$f = 2$ produced a herringbone state
that transitioned under an applied drive into an aligned 
ferromagnetic state where the particles move
in 1D channels \cite{C}. 
For weaker substrates a triangular lattice formed 
and the depinning was elastic,
while for intermediate substrate strengths
the depinning was plastic and the herringbone state 
depinned into a fluctuating state that reordered at higher drives.  
There was a strong increase in the critical 
depinning force between the elastic and plastic depinning regimes
similar to the peak effect phenomenon found at the
transition between elastic and plastic depinning for vortices
in type-II superconductors.  
For $f = 3$ the pinned state was always triangular and for weak substrates
the particles depinned into either a moving crystal or a moving smectic state. 
At these low fillings
it was also shown that for weak substrates 
the critical depinning force $F_{c} \propto A^2$,
as expected for elastic depinning, 
while for strong substrates, $F_{c} \propto A$,
as expected for single
particle depinning. 

\begin{figure}
\includegraphics[width=\columnwidth]{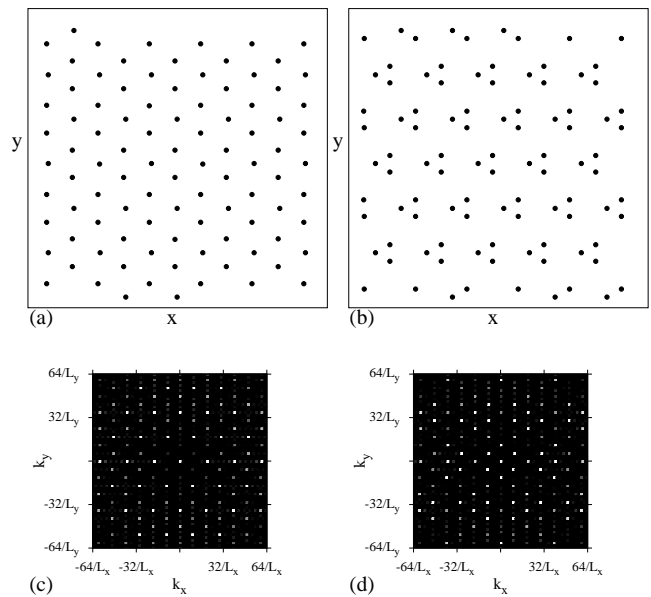}
\caption{
(a, b) The particle locations (black dots) for a triangular substrate
at a filling of $f =3$ for different substrate strengths of 
(a) $A = 1.0$ and (b) $A=3.0$.  
In (b) the trimer states that form at each trap
are aligned. 
(c) The structure factor $S(k)$ corresponding to panel (a) at $A=1.0$. 
(d) $S(k)$ corresponding to panel (b) at $A=3.0$
shows long range order and some features
at larger $k$ values corresponding to the smaller real space length
scale of the trimers.   
}
\label{fig1}
\end{figure}

We now consider several features at $f = 3$ that were not 
reported in the earlier studies, including the velocity-force curves 
and an additional dynamical phase.
Figures~\ref{fig1}(a) and \ref{fig1}(b) show the particle configurations at 
$f = 3$ for $A = 1.0$ and $A=3.0$, 
respectively, while
Figs.~\ref{fig1}(c) and \ref{fig1}(d) show the corresponding structure 
factor $S(k)$. 
For $A = 1.0$ in Fig.~\ref{fig1}(a,c), a triangular
lattice forms with one dominant length scale 
that is the distance between the
particles. 
For $A = 3.0$ in Fig.~\ref{fig1}(b,d),
each substrate minimum clearly captures three particles
and the structure factor has the signature of two length scales.
There is a well defined hexagonal pattern of maxima in $S(k)$ surrounding the
origin in Fig.~\ref{fig1}(d) that was not present in Fig.~\ref{fig1}(c) 
which is associated with
the longer length scale of the substrate minima locations.
The hexagonal void arrangement at 
larger $k$ values in Fig.~\ref{fig1}(d) is associated
with the smaller length scale of the 
trimer colloid arrangement within each substrate minimum.
In general, as $A$ increases further, the structure of the lattice remains
the same as shown in Fig.~\ref{fig1}(c,d)
except that the
trimers gradually reduce in size.  

\begin{figure}
\includegraphics[width=\columnwidth]{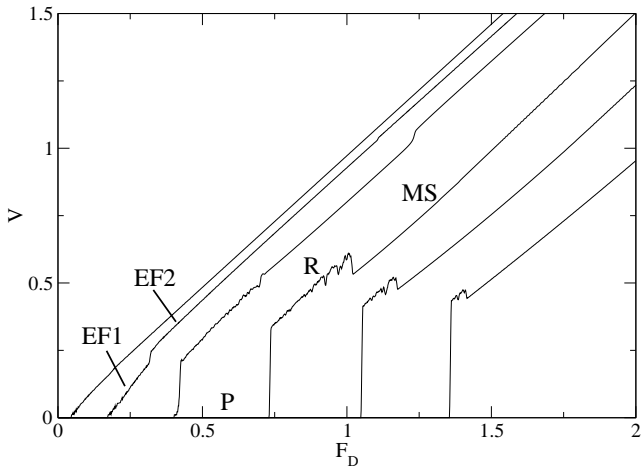}
\caption{
The average velocity 
$V$ vs external drive $F_{D}$ for the $f = 3$ system in Fig.~\ref{fig1}
for $A = 1$, 2, 3, 4, 5, and 6, from left to right. 
For $A = 4.0$ the three different phases
are labeled for the pinned (P), random flow (R), and moving smectic (MS) phases.
For the $A = 2.0$ curve there is no random phase but two different elastic flow
phases $EF1$ and $EF2$ occur which are labeled with lines.    
}
\label{fig2}
\end{figure}

\begin{figure}
\includegraphics[width=\columnwidth]{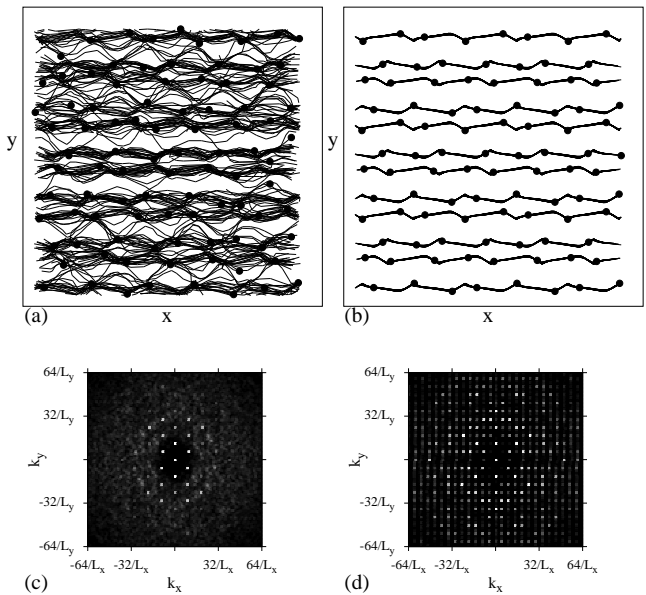}
\caption{
(a) 
Particle positions (dots) and trajectories (lines) over a fixed
time interval
for the random phase at $F_D=0.9$ and $A = 4.0$ 
for the $f=3$ sample from Fig.~\ref{fig2}.
The trajectories are rapidly changing with time
and mixing of the particles occurs. 
(c) The corresponding $S(k)$ 
shows sixfold peaks at small $k$ due to the overall triangular
ordering of the particles on the substrate as well as a ring structure 
at larger $k$ due to the liquidlike
nature of the particle positions. 
(b) Particle positions (dots) and trajectories (lines) over a fixed
time interval
in the moving smectic phase at $F_D=1.5$
for the same sample,
where the particles follow well-defined winding quasi-1D paths. 
(d) The corresponding $S(k)$ 
shows more ordering, 
including more pronounced peaks along $k_{y}$ that are consistent with
a smectic phase.    
}
\label{fig3}
\end{figure}

In Fig.~\ref{fig2} we plot the average particle velocity  $V$
versus the external drive $F_{D}$
for $A = 1$, 2, 3, 4, 5, and $6$. 
For the $A = 4.0$ curve
we have labeled the different dynamical phases. 
The pinned phase (P) 
transitions into a rapidly fluctuating phase we term the random phase
(R). 
The random phase is characterized by 
rapid mixing of the particles,
as illustrated in Fig.~\ref{fig3}(a) for $A = 4.0$ at $F_D=0.9$,   
where the particle trajectories show significant excursions
in the direction transverse to the drive
and are continuously changing with time. 
In Fig.~\ref{fig3}(b), the corresponding $S(k)$ taken from
a snapshot of the particle positions in the random phase
shows features of an anisotropic liquid.
There is triangular ordering in
$S(k)$ at smaller $k$ values 
due to the underlying triangular substrate; 
however, the higher order peaks of Fig.~\ref{fig1}(d) that indicated long range
order in the pinned state are replaced in Fig.~\ref{fig3}(b) by a
ring structure characteristic of a liquid. 
We also find
a larger number of peaks along $k_{y}$ compared to $k_{x}$ 
due to the anisotropy induced by the $x$-direction drive.
For higher drives the system transitions into a state with no
transverse diffusion, as illustrated in
Fig.~\ref{fig3}(b,d) for a sample with $A=4.0$ at $F_{D} = 1.5$. 
In this state, which we term the moving smectic (MS) phase,
the motion is confined to 
winding quasi-1D flows as shown in Fig.~\ref{fig3}(b). 
The plot of $S(k)$ in Fig.~\ref{fig3}(d) indicates that the MS state has stronger
partial ordering than the random phase but that there is still a tendency for 
the system to be more ordered along the $k_{y}$ direction. 

The R to MS transition is
correlated with a decrease in $V$ with increasing $F_{D}$, 
producing
a negative $dV/dF_{D}$ signature known as
negative differential conductivity (NDC).
Simulations and experiments on superconducting 
vortices moving over periodic pinning
sites have revealed NDC
at transitions from random
or turbulent flows to ordered 1D flows \cite{Nori}. 
The NDC in the vortex system is much more pronounced than the NDC we
observe in Fig.~\ref{fig2}, and this is likely due to the difference in the type
of substrate used in the two systems.
For the vortex system, artificially fabricated pinning sites produce a
short range muffin-tin potential that permits some vortices to sit 
completely outside of the pinning sites in the flat interstitial regions.
In the random phase, a large number of vortices move through the
interstitial areas and do not interact directly with the pins, but at the
transition to 1D flow, these vortices suddenly fall into flowing channels that
pass through the pinning sites, abruptly increasing the effective amount of
drag exerted by the pinning sites on the vortices.
In the sinusoidal triangular substrate we consider here, 
there are no interstitial regions and the moving
particles are always experiencing
some drag from the pinning.
Additionally, in the vortex system NDC 
was only associated with fillings just above $f=1$, the first matching field
\cite{Nori}, while in Fig.~\ref{fig2} we find no
NDC until $f  \geq 3$.  
We note that the NDC for the triangular substrate
is much more robust than NDC in the vortex system since it
appears for a much wider range of fillings, including incommensurate
fillings. 
In a recent vortex experiment \cite{Avci} NDC did not occur
until $f > 2.5$;
however, this effect could not be reproduced in
vortex simulations using only muffin-tin type pinning potentials.
This could mean that for the particular system considered in Ref.~\cite{Avci},
despite the fact that the pinning sites are localized, longer range 
interactions between the pins or distortions of the substrate may have 
caused the substrate to behave more like the sinusoidal type of substrate
we consider here rather than like a muffin-tin potential.

\begin{figure}
\includegraphics[width=\columnwidth]{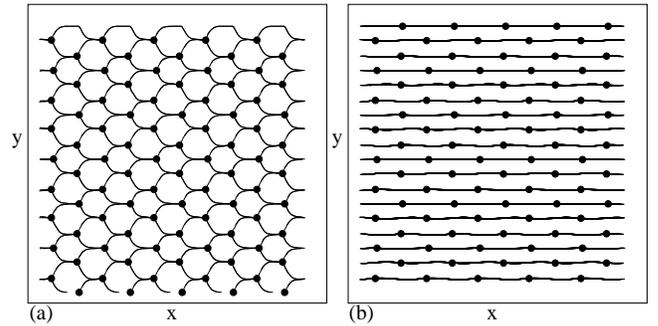}
\caption{
Particle positions (dots) and trajectories (lines) over a fixed time
interval in
the elastic flow phases from the $f=3$ sample in Fig.~\ref{fig2} at $A = 2.0$. 
(a) In phase EF1 at $F_D=0.25$, the particles form a triangular lattice and
flow through winding channels but there is no particle mixing.
(b) In phase EF2 at $F_D=0.5$, 
the trajectories become straight. The transition between these two phases 
appears as a kink in $V$ near $F_D=0.32$ in Fig.~\ref{fig2}.
}
\label{fig4}
\end{figure}

For $A>4.0$ we find the same three phases described for $A=4.0$, with the
transitions between phases shifting to higher $F_D$ with increasing $A$.
In Fig.~\ref{fig2} we identify two distinct phases observed for a weaker 
substrate with $A = 2.0$. 
Here the random phase is absent and the system depins 
elastically, with no sharp jump in $V$ at depinning as found for $A>2.0$ when
the flow above depinning is plastic.
Even though the depinning transition for $A=2.0$ is elastic, we still find
transitions between distinct dynamical phases that can be identified through
signatures in the velocity-force curve.  In Fig.~\ref{fig2} we mark the two elastic
flow phases EF1 and EF2 that appear on either side of a discontinuity in
$dV/dF_D$ centered at $F_D=0.32$.
In both phases the particle lattice is triangular and has an $S(k)$ similar
to that shown in Fig.~\ref{fig1}(b).
The EF1 and EF2 phases also appear for $A=3.0$ with the kink in $V$ shifted
up to higher $F_D$.
In Fig.~\ref{fig4}(a) we plot the particle
trajectories in phase EF1 for the $A=2.0$ system.  The triangular particle
lattice moves in a zig-zag pattern in order to avoid passing over the
potential maxima of the triangular substrate.  At higher drives the
particles no longer avoid the potential maxima and enter phase EF2 where they
move in the direction of drive, as illustrated in Fig.~\ref{fig4}(b).

\begin{figure}
\includegraphics[width=\columnwidth]{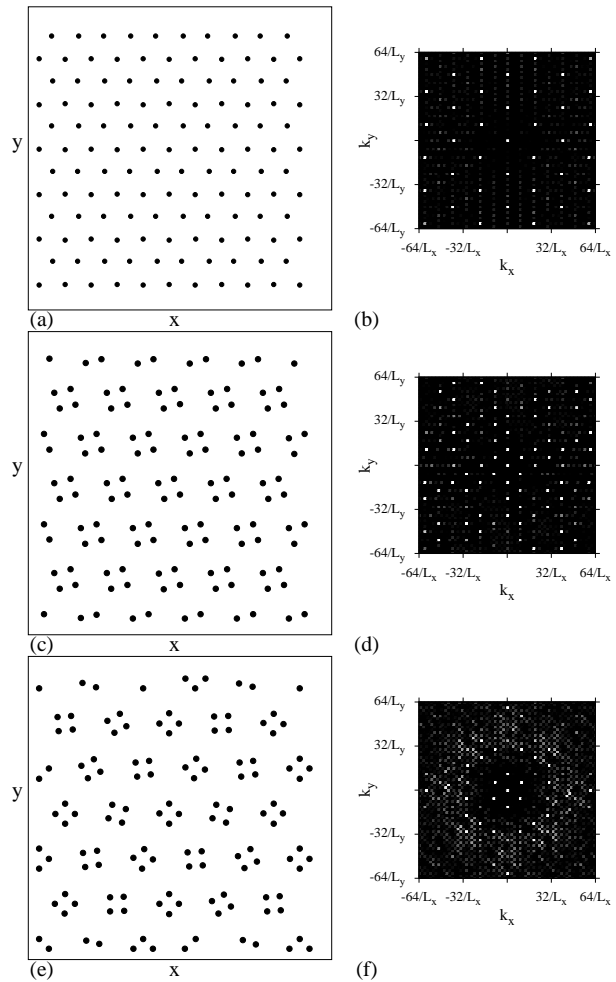}
\caption{ (a,c,e) The particle configurations and (b,d,f) 
corresponding $S(k)$ for samples with $f = 4$.
(a,b) At $A = 1.0$ the particles form a triangular lattice. 
(c,d) At $A = 8.0$ a quadrimer state forms. 
The quadrimers have a ferromagnetic alignment.
(e,f) At $A = 11.0$ the quadrimer state has 
no orientational ordering, producing
ring structures in $S(k)$.
}
\label{fig5}
\end{figure}

For $f = 4$, when the substrate is weak a triangular lattice of
particles forms, as illustrated in Fig.~\ref{fig5}(a,b) for a sample with $A=1.0$.
Weak substrates still affect the lattice by breaking rotational symmetry
and causing the particle lattice to preferentially orient 
in a direction determined by the filling factor.
At $f=4$, one of the symmetry axes of the particle
lattice is aligned with the $x$ axis as in Fig.~\ref{fig5}(a), while
in Fig.~\ref{fig1}(a) at $f=3$ it was aligned with the $y$ axis.
As the substrate strength increases for $f=4$, quadrimer arrangements of
the particles form in each substrate minimum, and the orientation of neighboring
quadrimers varies with $A$.
For $A = 8.0$, Fig.~\ref{fig5}(c) shows that all the quadrimers are aligned in
the same direction.
The corresponding $S(k)$ in Fig.~\ref{fig5}(d) shows sixfold peaks at small $k$
from the triangular substrate, while at larger $k$ a fourfold void structure
appears due to the square ordering of each quadrimer.
At $A = 11$ in Fig.~\ref{fig5}(e), the individual quadrimers remain intact but
their long range ferromagnetic orientational
ordering is lost,
producing a ring structure in $S(k)$ as shown in Fig.~\ref{fig5}(f).
We have tried several different methods of preparing the initial
configuration for $f=4$ and $A=11.0$ but have not found a ferromagnetically
ordered state.

The orientational ordering of dimer and trimer states was previously shown 
to arise due to quadrupole or higher pole moment interactions between
neighboring $n$-mers.
For quadrimers, it is likely that the higher pole moments play a more
important role.
If the pole moment favors
ferromagnetic alignment, than as $A$ increases the size
of the quadrimer decreases, reducing the pole moment responsible for the
orientational ordering. 
Previous studies found that as the substrate strength increases,
the temperature at which dimers and trimers lose their orientational
ordering drops due to the decreasing effective pole moment. 
We note that
the configurations in Fig.~\ref{fig5} are all obtained at $T = 0.0$.  
When we anneal the system
from a high temperature and slowly decrease the temperature 
to zero, we still obtain the same states illustrated in Fig.~\ref{fig5},
suggesting that the energy of the ferromagnetic state must
be only slightly smaller than that of the rotationally disordered state. 
As $A$ increases, the picture of rigid quadrimers begins to
break down, as shown in Fig.~\ref{fig5}(e) where the quadrimers become slightly
distorted.
This could further decrease the contributions of different pole moments to
the orientational ordering or may even lead to
competing interactions between neighboring quadrimers,
resulting in a ground state with ordering produced by very long range
interactions, as found in certain spin ice systems.
For $A>11.0$ we find
states similar to those shown in Fig.~\ref{fig5}(e,f) 
with increasing distortion in the square 
ordering, until for high enough $A$ the distortion becomes strong enough to
permit some particles to sit at the center of a substrate minimum.
The dynamic phases for $f=4$ are similar to those illustrated in Fig.~\ref{fig2}
for the $f=3$ system, 
with elastic depinning for low $A$ crossing over to plastic depinning at
higher $A$, followed by a dynamical reordering transition at higher $F_D$.

\begin{figure}
\includegraphics[width=\columnwidth]{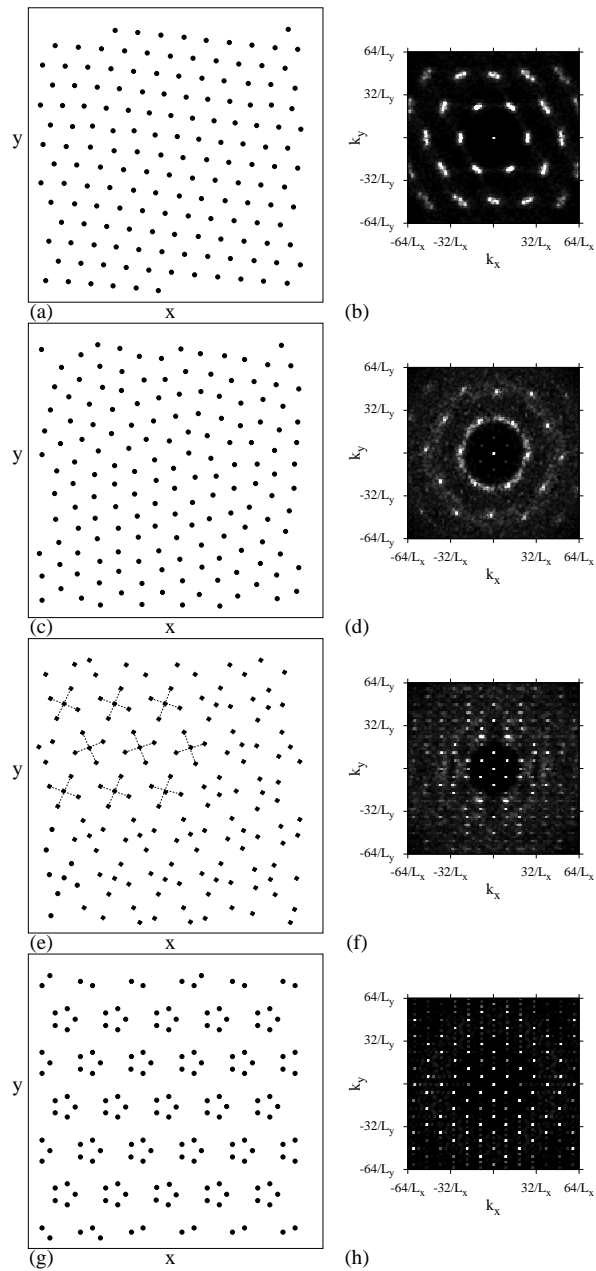}
\caption{
(a,c,e,g) The particle configurations and 
(b,d,f,h) corresponding $S(k)$ for samples with $f =5$.
(a,b) At $A = 0.5$ the particles have triangular ordering that 
is not commensurate
with the substrate, as indicated by the smearing of the peaks in $S(k)$.
(c,d) At $A = 1.75$ the system is disordered. 
(e,f) At $A = 4.1$ the system forms what we term a jack state 
with one particle located at the center of each potential minimum 
surrounded by four outer particles in a square configuration. 
In large regions the squares form a herringbone structure, as indicated
by the dashed lines; however,
some localized disorder is present in the sample.
(g,h) 
At $A=11.0$, an aligned pentagon state with long range orientational order
forms.
For $A>11.0$ the same structure persists but the pentagons shrink in size.
}
\label{fig6}
\end{figure}

For $f = 5$ at $F_D=0$, as a function of $A$ we find 
disordered phases interspersed among ordered phases.
This is also the first filling at which
the $n$-mer assumption breaks down 
and a shell structure of the particles in each substrate minimum
begins to form, in contrast to the case of isolated trapped
clusters of 
repulsively interacting particles,
which first develop equilibrium shell structure
at $f=6$ \cite{Peeters1,Lin,Even,bolton,buzdin,lai,grzybowski} but have
a metastable state with one particle in the center for $f=5$
\cite{campbell}.
In Fig.~\ref{fig6}(a) for $A=0.5$ at $f=5$, 
a triangular lattice containing a small tilt
distortion forms.  The smearing of the corresponding $S(k)$ in Fig.~\ref{fig6}(b)
indicates that for weak substrates at this filling, the 
particle lattice is not commensurate with the substrate.
At $A = 1.75$, shown in Fig.~\ref{fig1}(c,d), the lattice is disordered
as indicated by the ring pattern in $S(k)$.
When we vary the initialization protocol, 
we always observe a disordered lattice
for $0.9 \leq A \leq 2.0$. 
For $2.0 < A < 7.0$ the system orders again       
as shown in Fig.~\ref{fig6}(e,f) for $A = 5.0$. 
Each substrate minimum contains what we term a jack pattern
consisting of one particle located at the center
of the minimum surrounded by four particles in a square arrangement.
As indicated by the dashed lines in Fig.~\ref{fig6}(e), the jacks are tilted
at $+20^\circ$ or $-20^\circ$ to the $x$ axis in every other row,
producing a herringbone structure.
The ordering is not complete as there are local regions where the jack
structure breaks down, producing some smearing in $S(k)$ as plotted in 
Fig.~\ref{fig6}(f).
The structure can be viewed as the 
beginning of a shell ordering where, in each substrate minimum,
the first shell contains a single particle and the next
shell contains four particles.  
For $ A > 7.0$ the interactions between particles in neighboring substrate
minima are reduced and the pentagon ground state structure expected for
particles in an isolated trap emerges.  The pentagons exhibit ferromagnetic
order and are all aligned in the same direction, as illustrated in
Fig.~\ref{fig6}(g,h) for $A=11.0$.
For increasing $A$ the pentagons become smaller; however, 
we observe no further change in the structure.

\begin{figure}
\includegraphics[width=\columnwidth]{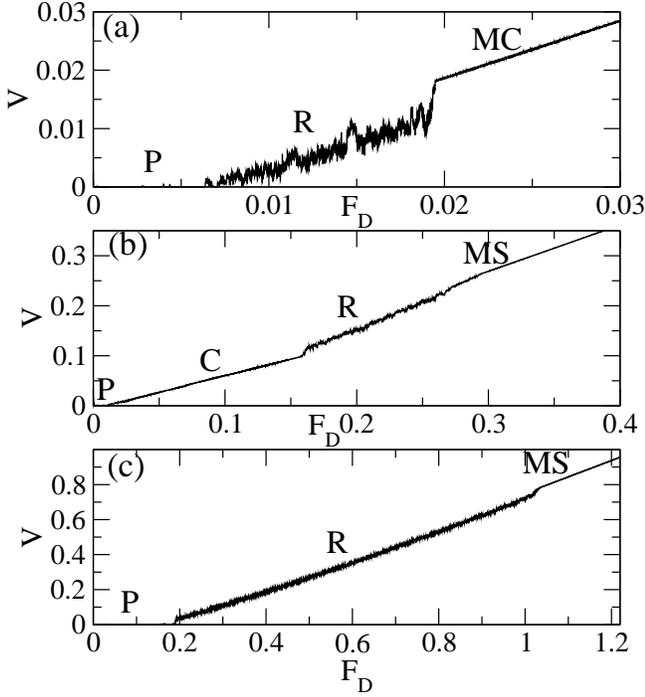}
\caption{
The velocity $V$ vs force $F_D$ curves for 
samples with $f = 5$. (a) At $A = 2.15$ we observe a pinned 
phase (P), a random phase (R), and a moving crystal phase (MC). 
(b) At $A = 4.1$ the system depins into a moving channel phase (C) 
illustrated in Fig.~\ref{fig8}(a). 
This is followed by the R phase and the moving smectic phase (MS).
(c) At $A = 7.2$ the C phase is absent and the system transitions
directly from the R phase to the MS phase.
}
\label{fig7}
\end{figure}

\begin{figure}
\includegraphics[width=\columnwidth]{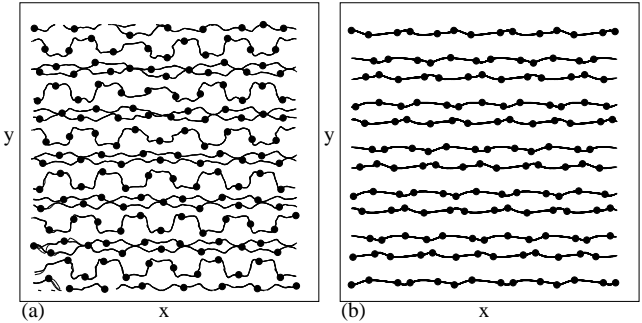}
\caption{
Particle positions (dots) and trajectories (lines) over a fixed time
interval for the $f=5$ sample.
(a) The moving channel (C) phase at $A = 4.1$ and $F_D=0.05$. 
(b) The moving smectic (MS) phase at $A = 7.2$ and $F_D=1.1$.  
}
\label{fig8}
\end{figure}

We can relate the different pinned structures at $f=5$
to features in the velocity
force curves and to dynamic phases. 
For $A < 1.5$ there is 
an elastic depinning into a moving triangular lattice. For $1.5 \leq A < 2.4$
the disordered phase depins into a fluctuating random phase similar
to that observed for $f = 3$, 
and at higher drives the system can reorder into
a moving crystal (MC) phase. 
In Fig.~\ref{fig7}(a) we plot $V$ versus $F_{D}$ for a sample with
$A = 2.15$. 
The random flow phase is characterized by large
fluctuations, and is followed by a sharp transition 
near $F_{D} = 0.034$ into a MC phase with small fluctuations. 
The transition
between these two phases can also be detected by analyzing the noise
fluctuations of the velocity response. 
In the random phase the fluctuations
are characterized by a broad band noise signal 
with $1/f^{\alpha}$ characteristics where
$\alpha = 1.5$ to $2.0$, while in the MC phase there is a narrow
band signal with a characteristic frequency that increases with increasing 
$F_{D}$, similar to what has been observed for dynamically reordered
phases of superconducting vortices moving over periodic substrates
\cite{Nori}. 
For $2.5 < A < 3.3$ we find no dynamical reordering and
the particle arrangement remains disordered up to the highest 
drives we considered; however, it is possible that a transition into a
moving ordered state could occur at very high drives.
For $3.1 < A < 6.1$, where the pinned system is in the ordered jack state, 
the depinning is elastic and the particles depin into the 
ordered winding channel flow phase (C) illustrated in Fig.~\ref{fig8}(a). 
As $F_D$ increases,
there is a crossover to a random phase followed by 
another transition at high drives into a moving smectic phase (MS) of the type
shown in Fig.~\ref{fig8}(b). 
The velocity signatures associated with these transitions appear
in Fig.~\ref{fig7}(b) for a sample with $A=4.1$.
The C phase is associated with small velocity fluctuations and
is followed first by the strongly fluctuating R phase and then
by the MS phase at higher $F_D$.
For $A \geq 6.0$ the 
C phase is lost and the sample depins
from the ordered pentagon state into a random flow
state, as illustrated in Fig.~\ref{fig7}(c) for a sample with $A=7.2$. 
This is followed by the formation of a moving smectic state at higher $F_D$.
As $A$ increases, the extent of both the pinned region and the 
random fluctuating state increases. 

\begin{figure}
\includegraphics[width=\columnwidth]{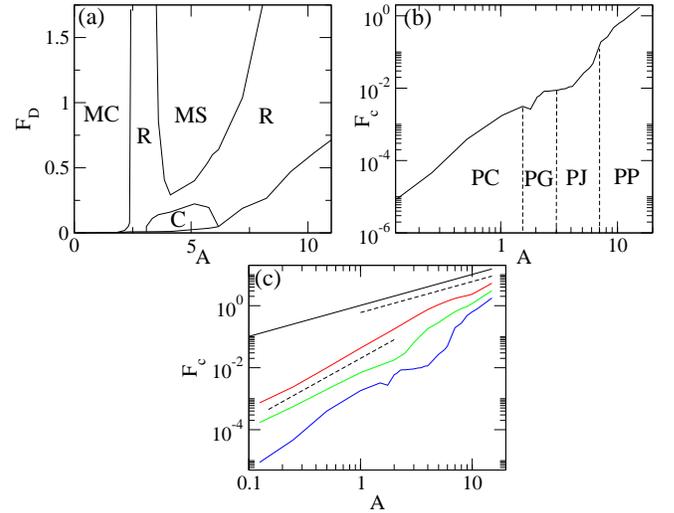}
\caption{
(a) The dynamical phase diagram $F_{D}$ vs $A$ for 
a sample with $f = 5$.  Regions where
the P, C, MC, R, and MS phases occur are marked. 
(b) The depinning curve $F_c$ from panel (a) plotted on a log-log scale.
Dashed lines mark the transitions between the
pinned triangular crystal (PC), pinned disordered or pinned glass (PG), 
pinned jack (PJ), and pinned pentagon (PP) states.
(c) The critical depinning force $F_{c}$ vs $A$ 
for $f = 1$, 3, 4, and $5$, from top to bottom.
Upper dashed line: a linear fit $F_c \propto A$. 
Lower dashed line: a fit to $F_{c} \propto A^2$.
These curves indicate that, in general, the system 
exhibits collective depinning behavior.  
}
\label{fig9}
\end{figure}

By conducting
a series of simulations, we map the dynamic phase diagram shown
in Fig.~\ref{fig9}(a). 
The disordered R phase separates the moving crystal and moving smectic 
phases.
There is also a domelike region where the winding channel C phase occurs. 
In Fig.~\ref{fig9}(b) we plot the depinning line on a log-log scale with
dashed lines marking the regions where the pinned phases illustrated in
Fig.~\ref{fig6} occur.
For $A < 0.9$, the depinning force $F_{c}$ monotonically
increases with increasing $A$. 
There is a small decrease in $F_c$ at the onset of 
the pinned disordered or pinned glass (PG) phase, followed by
a plateau in $F_c$. 
Near $A = 3.0$, $F_{c}$ begins to increase rapidly with
increasing $A$ in the pinned jack state, while in the
pinned pentagon state $F_c$ still increases with increasing $A$ but
with a reduced slope.
The behavior of $F_c$ 
near the onset of the pinned disordered phase is somewhat unusual since
it indicates that even though the substrate strength is increasing, the
depinning force does not increase. 
When the system is disordered, there are numerous dislocations present in the
particle structure that produce weak spots which flow first at the depinning
transition.  In contrast, in the depinning of a crystalline state,
there are no weak spots since the lattice structure is the same everywhere. 
This behavior is the opposite of the
so-called peak effect phenomenon found for vortices in type-II superconductors,
where a sudden increase in $F_{c}$ occurs when the vortex lattice 
becomes disordered. 
The peak effect is generally believed to arise
due to the softening of the vortex lattice 
when dislocations are present. 
In the softer lattice,
the vortices can more freely move to adjust to a random pinning potential
and maximize the pinning force. 
In our system, the pinning potential is not random,
so the crystalline states are more strongly pinned
than the disordered states. 
This has been demonstrated clearly in studies of commensurate-incommensurate
transitions for both vortices and colloids in 
periodic substrates, where crystalline states form
at integer values of $f$.  Near but not at commensuration, numerous defects
appear in the commensurate lattice, and these defects cause a reduction
in $F_c$.  As a result, $F_c$ passes through a series of peaks at integer
values of $f$ as the filling fraction is varied.
In the system we are considering here,
the $f = 5$ state is unusual since
even through the system is at an integer filling, we find
a regime where the pinned state is noncrystalline. 
This is in contrast to the lower integer fillings $f=1$, 2, 3, and 4,
where all the pinned states are ordered. 
This disordering effect and the
plateau or drop in $F_{c}$ at intermediate substrate strengths 
in the disordered pinned regions
is a general feature 
of $f \geq 5$ systems at integer fillings, and both features
becomes more prominent for higher fillings.
It may appear from Fig.~\ref{fig9}(b) that the plateau in $F_c$ 
occurs in the disordered region and not in the pinned jack state;
however, the plateau persists into a window of the pinned jack state
since the PJ state contains some defects.

In Fig.~\ref{fig9}(c) we plot $F_{c}$ 
vs $A$ for samples with $f = 1$, $3$, $4$, and $5$. The
dynamics and configurations of the $f = 2$ system 
were described in detail in 
previous work \cite{C}.
For $f = 1$ we find that $F_{c} \propto A$, 
as indicated by the upper dashed line.  This is expected in the limit
of single particle depinning and indicates that particle-particle interactions
are unimportant at depinning for this filling.
For the higher fillings $f>1$, we instead find
$F_{c} \propto A^{2}$ as indicated by the lower dashed line.
This is characteristic of
collective depinning transitions, where the interactions among the
particles within each minimum play an important role in
the depinning process.
In the limit of very strong substrates,
the depinning generally occurs when one particle is pushed near the saddle
point of the substrate by the other particles in the potential minimum.
This single particle suddenly escapes and triggers a cascade of escape 
events in the rest of the system.
At intermediate substrate strengths, the higher order fillings all 
generally show a change in the slope of $F_c$ versus $A$ associated
with a change in the $n$-mer structure.   
   
\section{Configurations and Dynamics for $6 \leq f  \leq 10$}       

\begin{figure}
\includegraphics[width=\columnwidth]{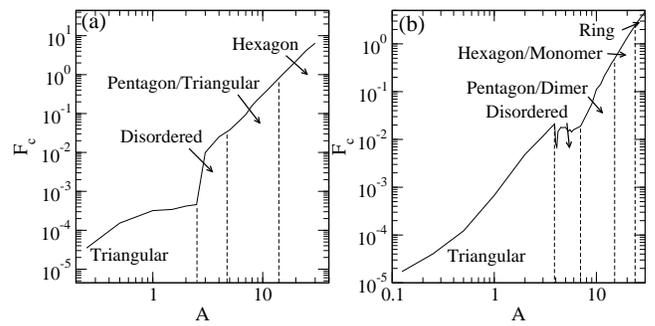}
\caption{
$F_{c}$ versus $A$. 
(a) At $f = 6$, a sharp increase in $F_{c}$ occurs at the
transition from a pinned triangular lattice to a disordered state. 
Dashed lines indicate regions where different pinned phases appear.
Triangular: triangular lattice; Disordered: disordered state; 
Pentagon/Triangular: a monomer at the center of each well surrounded
by five particles arranged in a pentagon; Hexagon: a hexagonal ring in
each substrate minimum.
(b) At $f = 7$, 
the triangular to disordered transition is associated with a small dip and
a plateau in $F_{c}$ with increasing $A$. 
Dashed lines indicate the regions where different pinned states appear.
Triangular: triangular lattice; Disordered: disordered state; 
Pentagon/Dimer: a composite pentagon-dimer arrangement in each
minimum; Hexagon/Monomer: a composite hexagon-monomer arrangement in
each minimum; Ring: a single ring of particles 
in each minimum found at large $A$.
}
\label{fig10}
\end{figure}

\begin{figure}
\includegraphics[width=\columnwidth]{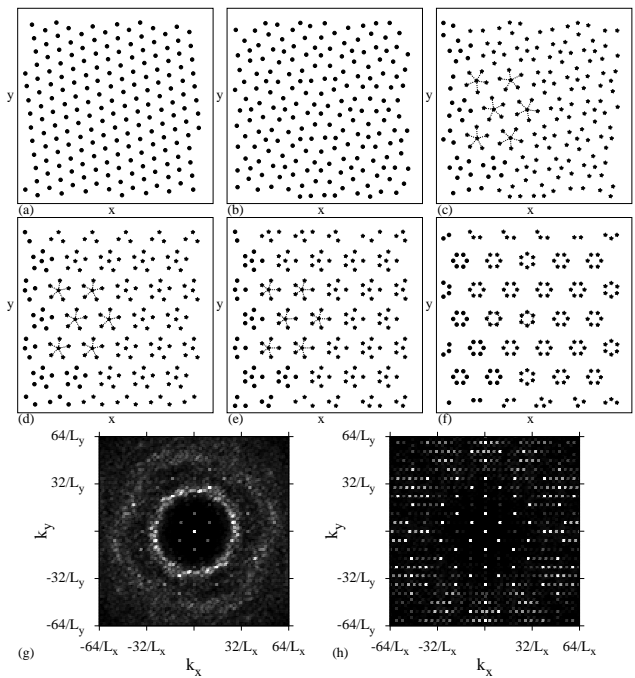}
\caption{
The particle positions after annealing for a sample with $f = 6$. In
panels (c,d,e), dashed lines indicate the orientation of the pentagons
in selected traps.
(a) At $A = 1.0$ and $F_D=0$ a triangular state forms. 
(b) At $A  = 2.75$ and $F_D=0$ a disordered state occurs. 
(c) At $A = 6.0$ and $F_D=0$ there is a pentagon/monomer state with no orientational
order.  
(d) At $A = 12.0$ and $F_D=0$ we find an aligned pentagon/monomer state.  
(e) At $A = 12.0$ and $F_D=0.45$, just below the depinning transition,
the pentagon/monomer state is polarized by the applied drive.  
(f) At $A=17.0$ and $F_D=0$, a hexagonal ring state with orientational
order forms.  Dashed lines indicate the two hexagon alignments
in representative traps.
(g) $S(k)$ for the disordered phase in panel (b). 
(h) $S(k)$ for the hexagonal ring state in panel (f).  
}
\label{fig11}
\end{figure}

In Fig.~\ref{fig10}(a) we plot $F_{c}$ versus $A$ for a sample with $f = 6$.
A large jump in $F_{c}$ occurs at $A=2.5$ at the transition from a pinned
triangular lattice, illustrated in Fig.~\ref{fig11}(a), 
to a disordered pinned state, illustrated in Fig.~\ref{fig11}(b).
For $2.5 < 5.5$, the system becomes disordered, as shown in 
Fig.~\ref{fig11}(b) for $ A = 2.75$, and has a ringlike $S(k)$ signature,
shown in Fig.~\ref{fig11}(g). 
In the disordered region 
the system depins into a plastic random flow state, 
while for $A < 2.5$ the system depins elastically 
into a MC state. 
At this filling, $F_c$ is depressed in the triangular state
since it is not possible for all the particles to simultaneously be in
a triangular lattice and sit at substrate minima locations, so that some
particles are instead located on substrate maxima.
When the substrate strength increases enough,
the particles are forced off of the substrate maxima and the
system disorders, allowing 
more particles to be located closer to substrate minima.     
For $5.5 <  A < 15$, a shell structure emerges.
The system forms a pentagonal/triangular lattice composite in which
one particle sits at each substrate minimum and is surrounded by five
particles in a pentagon arrangement, as illustrated in
Fig.~\ref{fig11}(c) for $A = 6.0$.  
In this case the ordering
is not ferromagnetic and the tilt of the pentagons varies from site to
site;
however, for $A > 8.0$ the
pentagon-monomer states become aligned, as shown in Fig.~\ref{fig11}(d) 
for $A = 10$. 
An applied drive can induce a polarization of the pentagons in the
driving direction, as illustrated in
Fig.~\ref{fig11}(e) for $A = 12.0$ under an $x$ direction drive
just before depinning. 
For $A > 15$ a ring state forms in which the center of the 
substrate minimum is empty and the particles form a hexagon. 
The rings are not all aligned; instead, two orientations coexist in the
sample as shown in Fig.~\ref{fig11}(f).  This produces twelve-fold modulations
in the structure factor as illustrated in Fig.~\ref{fig11}(h).
As $A$ increases further,
the ring structure persists but decreases in diameter. 
In general we find ring structures rather than shell structures
for $f \geq 6$ at large $A$.
The different phases are highlighted in Fig.~\ref{fig10}(a).  
 
\begin{figure}
\includegraphics[width=\columnwidth]{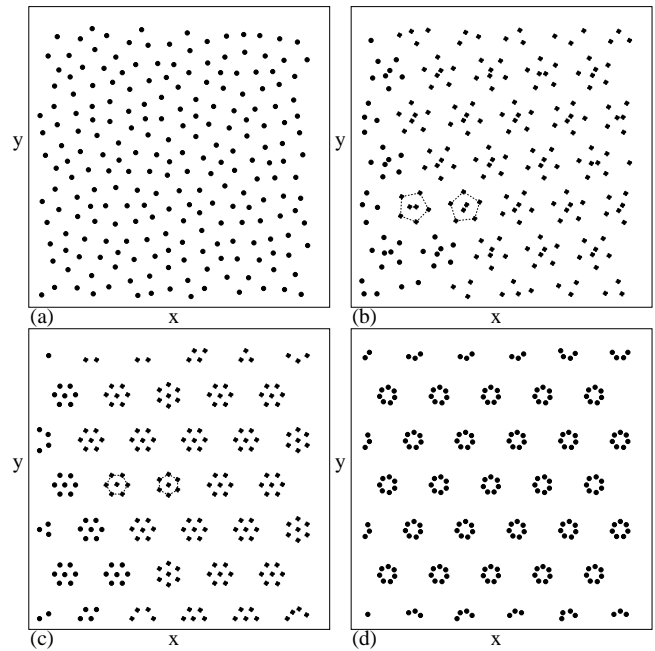}
\caption{
The particle positions after annealing for a sample with $f = 7$. 
(a) The disordered state at $A = 4.0$. 
(b)At $A = 12.0$ a pentagon/dimer composite appears in which
the dimers are partially aligned.  Dashed lines indicate two different
pentagon/dimer orientations.
(c) At $A = 20.0$ a hexagon/monomer composite forms. Dashed lines indicate
two different hexagon orientations.
(d) At $A=25.0$ a ring structure appears with seven particles per ring. 
}
\label{fig12}
\end{figure}

Figure~\ref{fig10}(b) shows that a similar set of pinned phases 
forms in a sample with $f = 7$. 
There is still a triangular lattice for small $A$; however, at $A=3.75$ when
the system enters the disordered phase, $F_c$ shows a small decrease followed
by a plateau up to the end of the disordered phase at $A=7.0$.
For $f=7$ 
the triangular lattice is more commensurate with the substrate and thus
better pinned than the triangular lattice at low $A$ in the $f=6$ system.
In the disordered phase, illustrated in Fig.~\ref{fig12}(a) for $A=4.0$,
$S(k)$ contains a ringlike structure similar to that in Fig.~\ref{fig11}(g).
For $7.0 < A < 16$ the system forms
a shell structure with an outer pentagon and two dimerized inner particles,
as shown  
in Fig.~\ref{fig12}(b) for $A = 12.0$. 
The pentagon/dimer structures are only locally aligned, leading to smearing
in the corresponding $S(k)$. 
Figure~\ref{fig12}(c) illustrates the configuration at $A = 20.0$ where  
an array of hexagons appears with a monomer at the center of each hexagon. 
The hexagons have one of two possible orientations, similar to the $f=6$
hexagon state shown in Fig.~\ref{fig11}(f).
For $A = 25$ shown in Fig.~\ref{fig12}(d), 
the shell ordering is replaced by a ring structure 
where each ring has seven particles and the ring radius 
decreases with increasing $A$. 
As also found for the $f = 6$ case,
it is possible for an external drive 
to orient the structures close to depinning. 

The dynamical phases for $f= 6$ and $f=7$ are very 
similar to those observed at $f = 5$. 
The system depins elastically at small $A$ in the triangular ordering 
regime and depins plastically in the disordered pinned regime.
A disordered flow regime extends to divergingly large drives for regions
of $A$ where a disordered pinned phase is present.
For values of $A$ above the disordered pinned phase, the strongly driven
system dynamically orders into a moving smectic phase,
and the driving force at which this transition occurs increases with
increasing $A$. 
For the higher fillings of $f=6$ and $f=7$, there are some additional 
features in the $V-F_{D}$ curves at high $A$ that did not appear at $f=5$.
For example, in the fluctuating phase, the system shows a coexistence
regime with some particles that remain pinned while other particles
move along 1D channels or troughs without rearranging the pinned particles. 
This feature becomes more noticeable at higher $f$.   

\begin{figure}
\includegraphics[width=\columnwidth]{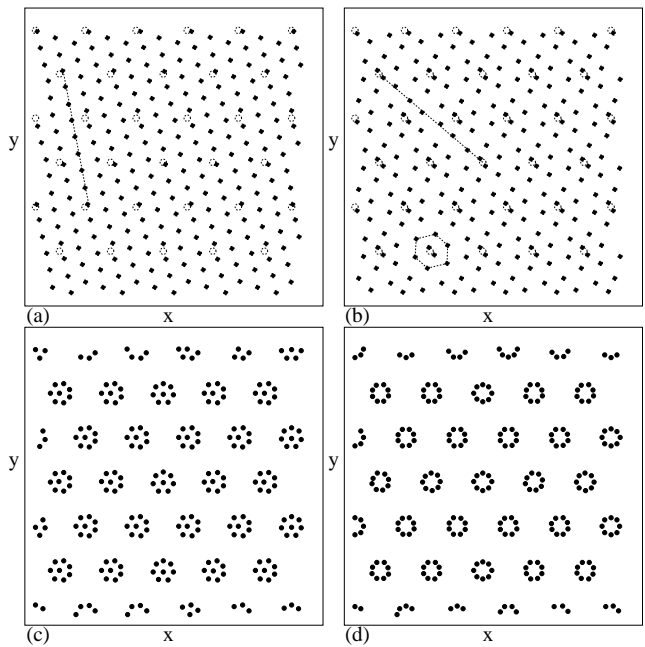}
\caption{
The particle positions after annealing for a sample with $f =8$.  In
(a,b), dashed circles indicate the centers of the substrate minima.
(a) At $A = 2.0$ a triangular lattice forms that
is rotated by $19.1^\circ$ with respect to the underlying substrate.  
The dashed line indicates
the direction along which 8 particles fit between consecutive substrate
minima.
(b) At $A = 5.0$, dimerization of a subset of the particles occurs
and the dimers are aligned approximately $-41^\circ$ from the $x$ axis.
The dashed line indicates the lattice alignment direction, and a representative
hexagon surrounding one dimer is shown.
(c) At $A = 20.0$, seven particles surround a single monomer in each minimum. 
(d) At $A = 26.0$ a ring state appears
with eight particles per ring.} 
\label{fig13}
\end{figure}

For $f = 8$ there are several new features in the dynamic phases 
due to the fact that there is no intermediate disordered pinned phase.
This results from a matching of the registry of the $f=8$ lattice with
the underlying substrate.
At $A=2.0$ in Fig.~\ref{fig13}(a), the particles form a triangular lattice
that is slightly anisotropic and 
rotated by $19.1^\circ$ with respect to the underlying substrate.
This orientation permits as many particles as possible to sit close to
substrate minima, and as indicated by the dashed line in Fig.~\ref{fig13}(a),
along one particle lattice symmetry direction, 
there are eight particles filling the space
between consecutive substrate minima.
This arrangement differs from the disordered lattice found at the
eighth matching field in a sample with a muffin-tin potential \cite{harada}.
As $A$ increases, each substrate minimum traps two particles that dimerize
and are surrounded by six other particles, as illustrated
in Fig.~\ref{fig13}(b).  The dimers are rotationally ordered,
and the overall particle lattice is rotated by $40.89^\circ$ with 
respect to the substrate, as indicated by the dashed 
line in Fig.~\ref{fig13}(b).
Orientationally ordered dimers did not form for $f=5$, 6, or 7.
For $A > 17.0$ the dimer state is lost and replaced by
a monomer at each substrate minimum surrounded by an
outer shell of seven particles, as shown in Fig.~\ref{fig13}(c) for $A=20.0$,
and for $A > 25.0$ a ring state with eight particles per ring occurs, 
as illustrated in Fig.~\ref{fig13}(d) for $A=26.0$. 

\begin{figure}
\includegraphics[width=\columnwidth]{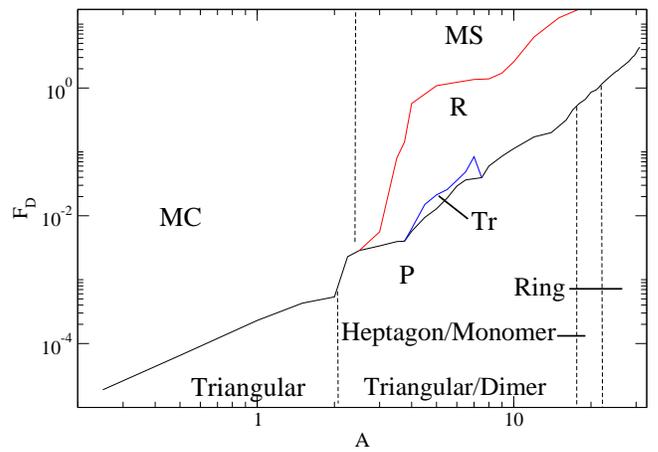}
\caption{
The dynamical phase diagram $F_D$ vs $A$ 
for a sample with $f=8$.
The pinned (P) phases include: a triangular lattice (Triangular), a triangular
lattice containing dimers (Triangular/Dimer), heptagons surrounding monomers
(Heptagon/Monomer), and a ring state (Ring).
There is a large increase in $F_{c}$ at the onset of the dimerized state. 
The transition flow region is labeled Tr.
The MC, MS, and R flow phases are also marked.
}   
\label{fig14}
\end{figure}

\begin{figure}
\includegraphics[width=\columnwidth]{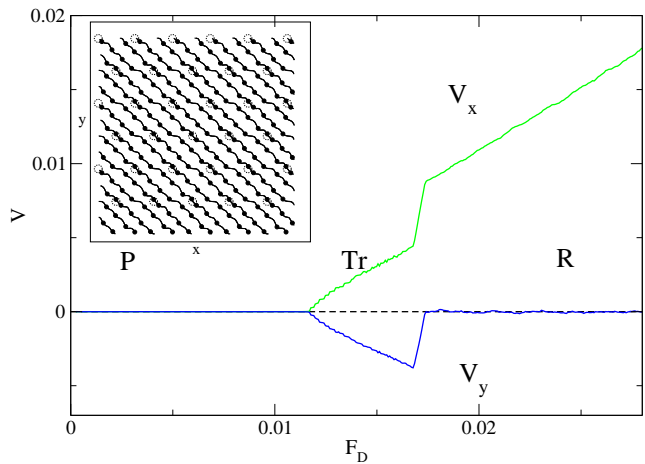}
\caption{
Main panel: $V_{x}$ (upper curve) and $V_{y}$ (lower curve)
vs $F_{D}$
for a sample with $f = 8$ at $A = 5.0$.  
P: pinned phase; Tr: transverse flow phase; R: random flow
phase with motion oriented along the $x$ direction only.  There is a
sharp transition between the Tr and R flow phases.
Inset: Particle positions (dots) and trajectories (lines) over a 
fixed time interval for the same sample in the 
transverse flow phase just above depinning at $F_D=0.0125$.
Dashed circles indicate the centers of the substrate minima.
The particles do not flow with the applied drive along the $x$-direction
but instead travel at $-40.89^\circ$ to the $x$ axis along one of the 
principal axes of the particle lattice. 
}   
\label{fig15}
\end{figure}

The lack of disorder for $f=8$ is clearly shown
in the dynamical phase diagram in Fig.~\ref{fig14}.
There are no disordered pinned phases, and the random flow phase (R) does
not extend out to large $F_D$ but is instead bounded on the high $F_D$
side by the MS flow state.
For $A \leq 2.0$, the triangular lattice depins elastically in the 
direction of drive into an MC state. 
There is a sharp increase in $F_c$ at the onset of the dimerized state,
and the MC flow is replaced by MS flow at higher drives.
For     
$ 4 \leq A < 6.0$, the system depins transverse to the driving
direction along one of the symmetry directions of the particle lattice
at $-40.89^\circ$ from the $x$ axis,
as illustrated in the inset of Fig.~\ref{fig15}.
As $F_{D}$ increases, there is a sharp transition out of the transverse
flow regime into 
a random flow phase with an average velocity oriented in the
driving direction, as shown in the main panel of Fig.~\ref{fig15} where we
plot $\langle V_x\rangle$ and $\langle V_y\rangle$ versus $F_D$.
In the transverse flow regime for $0.012 < F_D < 0.0175$,
$\langle V_y\rangle$ is negative and $\langle V_x\rangle$ is positive,
and both $|\langle V_x\rangle|$ and $\langle V_y\rangle$ increase
linearly with increasing drive until, at $F_D=0.0175$,
$\langle V_y\rangle$ drops to zero and $\langle V_x\rangle$ jumps up at
the transition to the random flow regime.
Note that the orientation of the pinned lattice is degenerate, so for samples
prepared with different random seeds, the particles would flow along
$-40.89^\circ$ in half the cases and along $+40.89^\circ$ in the other
half of the cases.
The transverse flow regime forms an intermediate state between the
pinned and random flow regimes.
Flows where the particles do not move in the direction
of drive but at an angle to the drive have previously been reported 
for colloids on triangular substrates
at a filling of $f= 2.0$ \cite{R}. 
In that study, the drive
was rotated $90^\circ$ 
relative to the substrate symmetry direction 
from the case considered here.
Dimerization of all the particles in the system created an orientational
degree of freedom that determined the flow direction of the colloids
just above depinning.
Numerical studies of
vortices in type-II superconductors with kagome and honeycomb pinning 
arrays also produced similar flows 
at angles to the driving direction in cases
where two vortices were located in the center of the kagome or 
honeycomb plaquettes, forming dimer states \cite{L}.  

\begin{figure}
\includegraphics[width=\columnwidth]{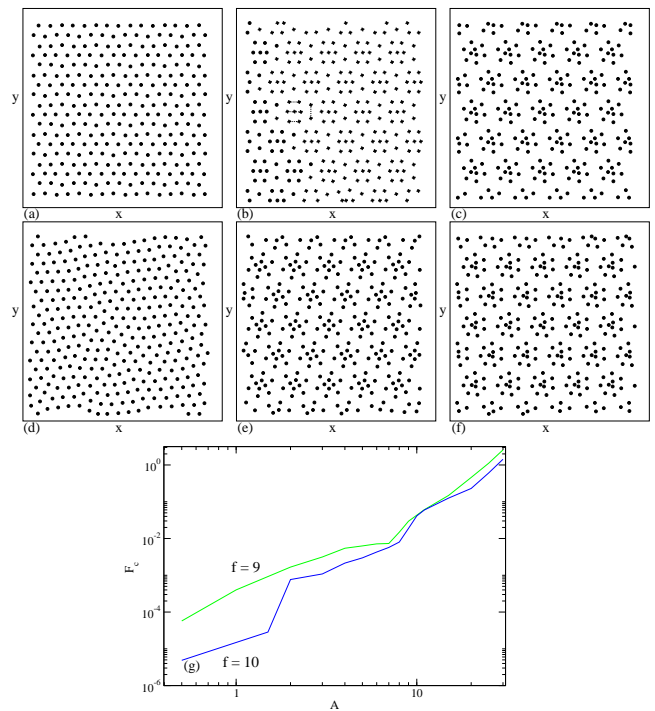}
\caption{(a-f) The particle positions after annealing.
(a) For $f = 9$ at $A =2.0$, a triangular lattice forms. 
(b) For $f = 9$ at $A = 7.0$, each substrate minimum contains a
linear trimer state aligned along the $x$ direction surrounded by
dimer states, as indicated by the dashed lines.
(c) For $f = 9$ at $A =15.0$, triangular trimers sit at each substrate
minimum surrounded by a hexagon of six particles.
(d) For $f = 10$ at $A = 1.0$, a partially ordered triangular state
occurs.
(e) For $f = 10$ at $A = 12.0$, aligned quadrimers form at each substrate
minimum surrounded by six particles.
(f) For $f = 10$ at $A = 15.0$, aligned triangular trimers appear that are
surrounded by seven particles.
(g) $F_{c}$  vs $A$ for $f = 9$ (upper curve) and 
$f = 10$ (lower curve).  
} 
\label{fig16}
\end{figure}

In samples with $f=9$, triangular configurations appear at low $A$, as
shown in Fig.~\ref{fig16}(a) for $A=2.0$.  This is the same configuration
found at the ninth matching field for vortices in triangular pinning
arrays in Ref.~\cite{harada}.
As $A$ increases, the particles align with the $x$ axis in an
unusual new structure illustrated in Fig.~\ref{fig16}(b) for $A=7.0$.
The system breaks into dimer and linear trimer states.  Each substrate
minimum contains a linear trimer flanked by two dimers, all oriented in
the $x$ direction, while between adjacent minima there are elongated dimers
oriented in the $y$ direction.
For $A = 15.0$, Fig.~\ref{fig16}(c) shows that a superlattice of aligned 
triangular trimers appears at each substrate minimum, surrounded by an outer
shell of six particles forming a hexagon. 
As $A$ increases further, we find a transition to a state with a monomer at
each substrate minimum surrounded by an eight-particle shell, followed by
a ring state at high $A$ (not shown).
For the parameters we considered, we never observed a state with
two dimerized particles at the centers of the substrate minima surrounded
by seven outer shell particles.  Either this state appears only for
an extremely narrow range of $A$, or it is simply too high in energy
to form at all.
In general we find that at the higher fillings, certain shell combinations 
are not observed.  These are often, but not always, associated with
incommensurate structures such as odd-even shells like the dimer-heptamer
structure for $f=9$. 

For $f=10$ at $A = 1.0$, Fig.~\ref{fig16}(d) indicates that
a partially ordered triangular lattice appears.  This is a floating
disordered solid that is only very weakly pinned by the substrate.
At $A=1.5$, the system transitions to a pinned disordered solid or
pinned glass state (not shown), where the last vestiges of triangular
ordering are lost.  For $A \approx 7$, the particles begin to develop
an incipient quadrimer structure around each substrate minimum.  The
quadrimers are still very extended and the lattice is only partially
localized by the substrate minima.
At $A=9.0$, the quadrimers become well localized within the minima and
the state illustrated in Fig.~\ref{fig16}(e) for $A=12.0$ emerges,
with aligned quadrimers surrounded by hexagons.
At $A=15.0$, aligned triangular trimers sit at each substrate minimum 
as illustrated in Fig.~\ref{fig16}(f).  This is similar to the state
shown in Fig.~\ref{fig16}(c) except each trimer is surrounded by seven
particles in the outer ring instead of six.
A similar set of orderings as those found for $f=9$ occurs for
$f=10$ as $A$ increases until a ring state forms at high $A$.
In Fig.~\ref{fig16}(g) we plot $F_c$ versus $A$ for $f=9$ and $f=10$.
For $f = 9$, the triangular lattice in Fig.~\ref{fig16}(a) 
transitions into the aligned linear trimer state in Fig.~\ref{fig16}(b)
at $A=7.0$, and there is a kink in $F_c$ associated with this transition.
For $f = 10$, the low-$A$, partially ordered lattice 
illustrated in Fig.~\ref{fig16}(d) has a very low $F_{c}$ since it is still
floating above the substrate.
A rapid increase in $F_c$ at $A=1.5$ is correlated with the transition
to a pinned disordered solid.  $F_c$ gradually increases as the quadrimer
structure gradually organizes, until at $A=9$, when the quadrimers become
well localized within the substrate maxima into the state
shown in Fig.~\ref{fig16}(e), $F_c$ kinks upward again.
Near $A=20$ there is another upward kink in $F_c$ at the formation of the
ring state, which depin as if they have no orientational degree of freedom.

\section{Higher Order Fillings and Discussion}
For higher order fillings $f>10$, we generally observe the same features
described for the $f > 5$ fillings, so we believe that 
we have identified the generic features
that arise in this type of system. 
For higher fillings
it is likely that instead of only two shells of particles, three or more
particle shell structures will form.
The orientational ordering of particles trapped in adjacent substrate
minima is also likely to become more fragile as the number of shells
increases and the outer shell becomes more and more circular.
Thermal effects on multiple shell structures
are beyond the scope of this work; however, we expect that the shell
structures would have multiple disordering transitions as a function
of temperature that are distinct from the transitions observed at lower
fillings in colloidal molecular crystals \cite{Reichhardt,Olson2}.
Another factor that could influence the ordering is the 
shape of the substrate minima.
Here we considered a sinusoidal substrate; however, with optical trapping, 
other forms of substrates could be created which may favor or disfavor 
certain shell structures from forming. 

\begin{figure}
\includegraphics[width=\columnwidth]{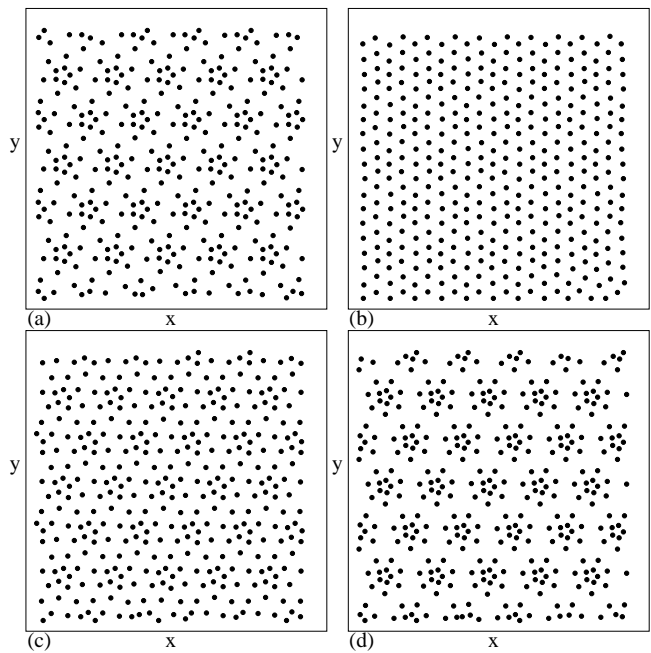}
\caption{
The particle configurations after annealing for high filling fractions, 
showing orientationally ordered states.
(a) For $f =11$ at $A = 14.0$, each substrate minimum contains
an inner shell of five particles and an outer shell of six particles.
(b) For $f = 12$ at $A = 1.0$, a triangular lattice appears. 
(c) For $f = 12$ at $A = 10.0$, there are four particles in the inner shells
surrounded by eight particles.
(d) For $f = 12$ at $A = 20$, there are five particles in the
inner shells surrounded by seven particles.   
} 
\label{fig17}
\end{figure}

In Fig.~\ref{fig17} we show some representative features 
of higher filling states.  
At $f=11$ and $A=14.0$ in Fig.~\ref{fig17}(a),
an aligned pentagonal inner ring forms at each substrate minimum
with an outer hexagonal ring.
For $f=12$, a triangular lattice appears for $A=1.0$ as illustrated
in Fig.~\ref{fig17}(b), an oriented structure with four particles in the
inner shell and nine in the outer shell forms at $A=10.0$ as shown in
Fig.~\ref{fig17}(c), and 
at $A = 20.0$ there are five particles in the inner shell and 
seven in the outer shell as seen in Fig.~\ref{fig17}(d).   

\section{Summary} 

We have examined the static configurations and driven dynamics of 
particles interacting with
a repulsive Yukawa potential, such as colloids, 
in the presence of a triangular substrate for varied integer fillings of up to
twelve particles per substrate minimum.
Under static conditions, we observe a rich variety of crystalline
structures for fillings that have not been previously 
reported for this model. 
We show that for fillings $f > 4$, shell structures can form in the
substrate minima
with well-defined numbers of particles in the inner and outer shells. 
These shell clusters can exhibit
orientational ordering across the sample.
As we vary the substrate strength for a fixed filling level,
we find that a series of different particle lattices occur
with different numbers of particles in the shells and different 
amounts of orientational ordering.
In the limit of strong substrates we observe a ring lattice. 
Several fillings exhibit disordered particle configurations; 
however, these fillings
can show a reentrant ordering behavior as the substrate strength
is increased.
Under an applied drive, 
the different particle orderings alter the critical depinning
force, producing either a strong increase in the depinning force 
or a plateau and even decrease in the depinning force at structural
transitions induced by increasing the substrate strength.
We also find a remarkably rich variety 
of dynamical phases that can be correlated with 
the structure of the static pinned configuration.
As the drive increases, we find
moving crystalline states and moving smectic states 
as well as disordered flow states.          
Our results are relevant to colloids on periodic substrates, 
vortices in nanostructured superconductors, and
other commensurate-incommensurate systems, 
including aspects of friction and nonequilibrium transport.  

\section{Acknowledgments}
This work was carried out under the auspices of the NNSA of the
U.S. DoE at LANL under 
Contract No.~DE-AC52-06NA25396.

\end{document}